\documentclass[11pt]{article}
\usepackage{jheppub}
\usepackage[latin1]{inputenc}
\usepackage[T1]{fontenc}
\usepackage{graphicx}
\usepackage{amsmath}
\usepackage{amsfonts}
\usepackage{amssymb}
\usepackage{latexsym}
\usepackage{slashed}
\usepackage{hyperref}
\usepackage{feynmp-auto}
\usepackage{mathtools}
\usepackage{subcaption}
\usepackage{tabularx}

\begin{document}

\title{Measuring neutrino dynamics in NMSSM with a right-handed sneutrino LSP at the ILC}
\author[1]{Yi Liu,}
\author[1,2]{Stefano Moretti}
\author[1,2]{and Harri Waltari}

\emailAdd{Yi.Liu@soton.ac.uk}
\emailAdd{S.Moretti@soton.ac.uk}
\emailAdd{H.Waltari@soton.ac.uk}

\affiliation[1]{School of Physics and Astronomy, University of Southampton, \\
Highfield, Southampton SO17 1BJ, UK}

\affiliation[2]{Particle Physics Department, STFC Rutherford Appleton Laboratory, \\
Chilton, Didcot, Oxon OX11 0QX, UK}

\date{\today}

\abstract{We study the possibility of measuring neutrino Yukawa couplings in the Next-to-Minimal Supersymmetric Standard Model  with right-handed neutrinos (NMSSMr) when the lightest right-handed sneutrino is the Dark Matter (DM) candidate, by exploiting a `dijet + dilepton + Missing Transverse Energy' (MET or $\slashed{E}_T$)  signature. We show that, contrary to the miminal realisation of Supersymmetry (SUSY), the MSSM, wherein the DM candidate is typically a much heavier (fermionic) neutralino state, this extended model of SUSY offers one with a much lighter (bosonic) state as DM that can then be produced at the next generation of $e^+e^-$ colliders with energies up to 500 GeV or so. The ensuing signal, energing from chargino pair production and subsequent decay, is extremely pure so it also affords one with the possibility of extracting the Yukawa parameters of the (s)neutrino sector. Altogether, our results serve the purpose of motivating searches for light DM signals at such machines, where the DM candidate can have a mass around the Electro-Weak (EW) scale.}     

\maketitle

\section{Introduction}

The Large Hadron Collider (LHC) experiments have so far shown good agreement with the predictions of the Standard Model (SM). Other types of experiments instead show us that the SM needs to be extended as neutrino oscillations require neutrinos to be massive \cite{Athanassopoulos:1997pv,Fukuda:1998mi,Aguilar:2001ty,Ahn:2002up,Abe:2011sj,An:2012eh}. Also the Cosmic Microwave Background (CMB) \cite{Ade:2013zuv} and galactic rotation curves \cite{Zwicky:1933gu,Rubin:1970zza} strongly support the idea that most of the mass of the Universe is in a form currently unknown to us, dubbed as Dark Matter (DM).

Neutrino oscillation experiments have measured the mixing angles of the Pontecorvo-Maki-Nakagawa-Sakata (PMNS) matrix \cite{Maki:1962mu} and the mass splittings between neutrinos, but they offer no information on the origin of neutrino masses. As the masses are several orders of magnitude smaller than any other fermion mass, it is expected that neutrino mass generation is based on some kind of a seesaw mechanism \cite{Minkowski:1977sc,Konetschny:1977bn,Mohapatra:1979ia,Magg:1980ut,Schechter:1980gr,Foot:1988aq}. The canonical example is Type-I seesaw, where one add heavy right-handed neutrinos to the SM particle spectrum, which then suppress the left-handed neutrino masses by a factor of $m_{D}/m_{N}$, where $m_{D}$ is the standard Dirac mass term and $m_{N}$ is the bare mass of the right-handed neutrino.

Supersymmetry (SUSY) is one of the most studied frameworks to construct Beyond the SM (BSM) theories. The Minimal Supersymmetric Standard Model (MSSM) still lacks a mechanism for neutrino mass generation, but it can be extended with various seesaw mechanisms. Supersymmetric models have the advantage that the superpartners of neutrinos, sneutrinos, may in some cases decay visibly at colliders and hence studying neutrino dynamics at colliders becomes possible \cite{Moretti:2019yln,Chakraborty:2020cpa}. However, current experimental constraints make it clear that non-minimal versions of SUSY are more suited to embed a neutrino mass generation mechanism \cite{Moretti:2019ulc}.

Adding a seesaw mechanism to models of SUSY gives also the option of non-standard DM candidates. Especially the right-handed sneutrino, when is the Lightest Supersymmetric Particle (LSP), has been of considerable interest over the years \cite{Asaka:2005cn,Gopalakrishna:2006kr,Lee:2007mt,Khalil:2011tb,Bandyopadhyay:2011qm,Belanger:2011rs,Basso:2012gz,Huitu:2012rd,Chatterjee:2014bva,Banerjee:2016uyt,Frank:2017tsm,DelleRose:2017uas,Banerjee:2018uut,Cao:2018iyk}. In the MSSM extended with Type-I seesaw right-handed sneutrinos lead to overabundance of the relic density of the CMB, unless there is significant mixing between the left- and right-handed sneutrinos \cite{ArkaniHamed:2000bq}. Adding a singlet to the model, (\textit{i.e.}, considering the Next-to-MSSM with right-handed neutrinos (NMSSMr)) allows a coupling between the heavy Higgses and sneutrinos, which can assist in the annihilation and lead to the correct relic abundance without the need for any left-right mixing in the sneutrino sector \cite{Cerdeno:2008ep,Cerdeno:2009dv}. 

We shall investigate here the possibility of extracting $e^+e^-$ collider signals and of estimating neutrino Yukawa couplings in the NMSSMr with a right-handed sneutrino as DM candidate. Since all superpartner decay chains end with this sneutrino, we cannot rely on sneutrino decays, instead we need to find visible decays that involve the sneutrino and  neutrino Yukawa couplings. Hence we need to study chargino decays to a lepton and a sneutrino. As the seesaw scale is rather low, the neutrino Yukawa couplings are tiny. Since the chargino can also decay to a virtual $W^\pm$ boson and a neutralino, the direct two-body decay of interest to us,  into a lepton-sneutrino pair,   is rare. However, the latter neutralino decay mode allows us to estimate the neutrino Yukawa couplings. Finally, the fact that right-handed sneutrinos can be rather light in the NMSSMr allows for DM signals emerging from the rare chargino decays to be sizable, particularly at future leptonic machines, where the SM background can efficiently be vetoed thanks to the high level of control on the final state kinematics. In the end, the signature that we will pursue is the one made up by a `dijet + dilepton + Missing Transverse Energy' (MET or ${\slashed{E}}_T$) system.

The plan of the paper is as follows. In the next section we describe our model. In the following one we illustrate the approach to be taken to extract the aforementioned rare chargino decay  ({\it i.e.}, $\tilde{\chi}^-\to l^-\tilde N$) and the numerical tools adopted to establish the significance of our signals. We then present our results for an $e^+e^-$ collider running at a Centre-of-Mass (CM) energy of 500 GeV and finally conclude. 

\section{NMSSM with right-handed neutrinos}

The NMSSM extends the MSSM with an additional gauge singlet chiral superfield $S$ \cite{Ellwanger:2009dp}. The NMSSM fixes the $\mu$ problem of MSSM by generating an effective $\mu$-term, but  it still inherits the defect that neutrinos are massless. By adding a singlet right-handed neutrino superfield $N$, we may introduce the Type-I seesaw mechanism to generate neutrino masses. The superpotential is given by \cite{Kitano:1999qb,Cerdeno:2008ep}

\begin{equation}
        W = W_{\rm NMSSM} + \lambda_N SNN + y_N H_2\cdot LN,
        \label{1}
    \end{equation}
    \begin{equation}
        W_{\rm NMSSM} = y_uH_2\cdot Qu + y_d H_1\cdot Qd + y_e H_1\cdot Le - \lambda SH_1\cdot H_2 + \frac{1}{3}\kappa S^3.
        \label{2}
    \end{equation} 

The flavour indices are omitted. As in the NMSSM, a $\mathbb{Z}_3$-symmetry is imposed in order to make the superpotential scale invariant. When this discrete symmetry is broken spontaneously by the Vacuum Expectation Values (VEVs) of the (pseudo)scalar fields, a potential domain wall problem arises. This problem can be solved like in the NMSSM, by assuming that non-renormalisable terms pick a preferred vacuum \cite{Abel:1995wk,Panagiotakopoulos:1998yw} or by supergravity corrections during inflation \cite{Mazumdar:2015dwd}\footnote{For an alternative formulation, called new Minimally-extended Supersymmetric Standard Model (nMSSM), where the domain wall (as well as the Peccei-Quinn axion) problem is solved by invoking a global discrete $R$-symmetry, see Ref.~\cite{Dedes:2000jp}.}.

The soft SUSY breaking terms are
\begin{equation}
    \begin{split}
     -\mathcal{L}_{soft} = & \ m^2_{\Tilde{Q}}|\Tilde{Q}|^2+m^2_{\Tilde{u}}|\Tilde{u}|^2+m^2_{\Tilde{d}}|\Tilde{d}|^2+m^2_{\Tilde{L}}|\Tilde{L}|^2+m^2_{\Tilde{e}}|\Tilde{e}|^2 +m^2_{\Tilde{N}}|\Tilde{N}|^2+m^2_S|S|^2\\&+m^2_{H_1}|H_1|^2 + m^2_{H_2}|H_2|^2  + M_1\Tilde{B}\Tilde{B}+M_2\Tilde{W}\Tilde{W}+M_3\Tilde{g}\Tilde{g} \\
     & +A_uY_uH_2\Tilde{Q}\Tilde{u} + A_dY_dH_1\Tilde{Q}\Tilde{d} + A_eY_eH_1\Tilde{L}\Tilde{e} + y_NA_{yN}\Tilde{L}H_2\Tilde{N} \\
     & -\lambda A_{\lambda}SH_1H_2 + \lambda_NA_{\lambda_N}SN^2+\frac{1}{3}\kappa A_{\kappa}S^3 +h.c.
    \end{split}
\end{equation}

Right-handed neutrino masses are generated when the scalar component of the singlet superfield $S$ gets a VEV, $\langle S \rangle=v_{s}$. The superpotential term $\lambda_N SNN$ in Eq.~(\ref{1}) leads to a Majorana mass term $M_N=2\lambda_N v_s$ so the right-handed neutrino masses are naturally at the Electro-Weak (EW) scale. The left-handed neutrino masses are obtained after the Higgs doublet fields acquire VEVs, $(v_1, v_2)=(\langle H_1 \rangle,\langle H_2 \rangle)$. The standard seesaw formula then gives $m_{\nu}=y_N^2 v_2^2 / M_N$. To get viable neutrino masses the neutrino Yukawa couplings $y_N$ have to be slightly smaller than the electron Yukawa coupling, $y_N \lesssim \mathcal{O}(10^{-6})$.

The left-hand sneutrino $\Tilde{\nu}_L$ and right-hand sneutrino $\Tilde{N}$ can be decomposed to CP-even (real) and CP-odd (imaginary) components:
\begin{equation}
    \Tilde{\nu}_L\equiv \frac{1}{\sqrt{2}}(\Tilde{\nu}_{L1}+i\Tilde{\nu}_{L2}), \qquad \Tilde{N}\equiv\frac{1}{\sqrt{2}}(\Tilde{N}_1+i\Tilde{N}_2).
\end{equation}
The sneutrino quadratic term is
\begin{equation}
    \frac{1}{2} \begin{array}{cccc} (\Tilde{\nu}_{L1},\Tilde{N}_1,\Tilde{\nu}_{L2},\Tilde{N}_2) 
    \end{array} \mathcal{M}^2_{\rm sneutrino}\left(\begin{array}{c}\title{\nu}_{L1}\\ \Tilde{N}_1\\ \Tilde{\nu}_{L2}\\ \Tilde{N}_2
    \end{array}\right).
    \label{5}
\end{equation}
The sneutrino mass matrix can be obtained from the quadratic terms in the scalar potential:
\begin{equation}
    \mathcal{M}_{\rm sneutrino}^2 =\left( \begin{array}{cccc}
          m_{L\bar{L}}^2 & \frac{m^2_{LR}+m^2_{L\bar{R}}+c.c}{2} & 0 & i\frac{m^2_{LR}-m^2_{L\bar{R}}-c.c}{2} \\ \frac{m^2_{LR}+m^2_{L\bar{R}}+c.c}{2} & m^2_{R\bar{R}}+M^2_{RR}+m^{2*}_{RR} & i\frac{m^2_{LR}-m^2_{L\bar{R}}-c.c}{2} & i(m^2_{RR}-m^{2*}_{RR}) \\ 0 & i\frac{m^2_{LR}-m^2_{L\bar{R}}-c.c}{2} & m^2_{L\bar{L}} & -\frac{m^2_{LR}+m^2_{L\bar{R}}+c.c}{2} \\ i\frac{m^2_{LR}-m^2_{L\bar{R}}-c.c}{2} & i(m^2_{RR}-m^{2*}_{RR} & -\frac{m^2_{LR}+m^2_{L\bar{R}}+c.c}{2} & m^2_{R\bar{R}}-M^2_{RR}-m^{2*}_{RR}
    \end{array}\right).
\end{equation}
The parameters are defined as follows:
\begin{equation}
\begin{split}
    & m^2_{L\bar{L}} \equiv m^2_{\Tilde{L}}+|y_Nv_2|^2 + {\rm D}-{\rm term}, \\
    & m^2_{LR} \equiv y_N(-\lambda v_sv_1)^\dagger + y_NA_Nv_2, \\
    & m^2_{L\bar{R}} \equiv y_Nv_2(-\lambda v_s)^\dagger, \\
    & m^2_{R\bar{R}} \equiv m^2_{\Tilde{N}}+|2\lambda_Nv_s|^2+|y_Nv_2|^2, \\
    & m^2_{RR} \equiv \lambda_N(A_{\lambda_N}v_s+(\kappa v_s^2-\lambda v_1v_2)^\dagger). \\
\end{split}
\label{7}
\end{equation}
Here, the $m^2_{\Tilde{L}}$, $m^2_{\Tilde{N}}$, $A_{\lambda_N}$ and $A_N$ are the soft SUSY breaking terms. Assuming thet there is no CP-violation, which means the sneutrino real part and imaginary part do not mix,  Eq.~(\ref{5}) can be simplified as:

\begin{equation}
\begin{split}
    & \frac{1}{2}(\Tilde{v}_{L1} \Tilde{N}_1) 
    \left( \begin{array}{cc} 
    m^2_{L\bar{L}} & m^2_{LR}+m^2_{L\bar{R}} \\ m^2_{LR}+m^2_{L\bar{R}} & m^2_{R\bar{R}}+2m^2_{RR}
    \end{array}\right) 
    \left( \begin{array}{c}
         \Tilde{v}_{L1} \\ \Tilde{N_1}
    \end{array}\right) + \\
    & \frac{1}{2}(\Tilde{v}_{L2} \Tilde{N}_2) 
    \left( \begin{array}{cc} 
    m^2_{L\bar{L}} & m^2_{LR}-m^2_{L\bar{R}} \\ m^2_{LR}-m^2_{L\bar{R}} & m^2_{R\bar{R}}-2m^2_{RR}
    \end{array}\right) 
    \left( \begin{array}{c}
         \Tilde{v}_{L2} \\ \Tilde{N_2}
    \end{array}\right).
\end{split}
\end{equation}

The mixing between left- and right-handed sneutrinos is determined by $m^2_{LR}$ and $m^2_{L\bar{R}}$. From Eq.~(\ref{7}), these two terms are proportional to the neutrino Yukawa coupling and therefore can be neglected. The mass difference between $\Tilde{N}_1$ and $\Tilde{N}_2$ is from the term $m^2_{RR}$. If $m^2_{RR}>0$, $\Tilde{N}_1$ is heavier than $\Tilde{N}_2$ and vice versa. In this case, the lighter right-handed sneutrino mass can be determined by $m^2_{R\bar{R}}-2m^2_{RR}$, which is defined by a set of parameters such as $m^2_{\Tilde{N}}$ and $\lambda_N$.

\subsection{Right-handed sneutrino as a dark matter candidate}

In the following we shall assume that the soft SUSY breaking masses for right-handed sneutrinos $m_{\tilde{N}}^{2}$ are the smallest of the SUSY breaking mass terms. Then there is a part of the parameter space where the LSP is the right-handed sneutrino. In the MSSM with right-handed neutrinos (MSSMr) it is not easy to satisfy the constraint from the relic density with a right-handed sneutrino LSP, since as a gauge singlet it does not annihilate efficiently enough. Enhancing the mixing between left- and right-handed sneutrinos is a way to remedy this \cite{ArkaniHamed:2000bq}.

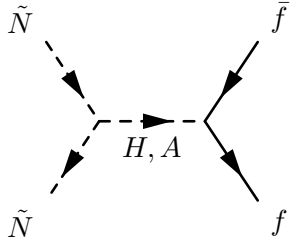
\begin{figure}
\begin{center}
\begin{fmffile}{annihilation}
  \begin{fmfgraph*}(100,60)
  \fmfleft{i1,i2}
  \fmfright{o1,o2}
  \fmf{scalar}{i2,v1,i1}
  \fmflabel{$\tilde{N}$}{i1}
  \fmflabel{$\tilde{N}$}{i2}
  \fmf{scalar,label=$H,A$}{v1,v2}
  \fmf{fermion}{o2,v2,o1}
  \fmflabel{$f$}{o1}
  \fmflabel{$\bar{f}$}{o2}
  \end{fmfgraph*}
\end{fmffile}
\end{center}
\caption{The dominant annihilation mechanism of sneutrino DM in the NMSSMr.}
\label{fig:relic}
\end{figure}

The NMSSM offers an additional method to enhance the annihilation cross section. The scalar potential has a term $\lambda\lambda_{N}H_{1}H_{2}\tilde{N}\tilde{N}$ which, after EW Symmetry Breaking (EWSB), creates a three-point coupling between the right-handed sneutrinos and Higgs bosons. The coupling between the sneutrinos and the heavy Higgses $H$ and $A$ is larger than that with the SM-like Higgs, 
so the sneutrino DM annihilates mostly via the heavy Higgs portal to third generation fermions (see Figure~\ref{fig:relic}), hence, 
the corresponding cross section mainly depends on $\lambda$, $\lambda_{N}$ and $m_{H,A}$. Further, in certain parts of the parameter space, coannihilations or resonant annihilation may alter the relic density largely.

The same Higgs-sneutrino couplings are mainly responsible for the effective sneutrino-nucleon interaction. In general the spin-independent direct detection cross sections are about one order of magnitude below the current experimental limit. Both constraints from the relic density and direct detection have been studied in the literature, most recently in \cite{Lopez-Fogliani:2021qpq,Kim:2021suj}.

\section{Finding the rare chargino decay}

We wish to estimate the neutrino Yukawa couplings in the presence of a right-handed sneutrino LSP. The decay $\tilde{\chi}^{\pm}\rightarrow \ell^{\pm}\tilde{N}$ arises from the neutrino Yukawa couplings. However, it competes with the decay $\tilde{\chi}^{\pm}\rightarrow W^{*\pm}\tilde{\chi}^{0}$, where the neutralino then decays to a right-handed sneutrino and some other particles. Even though this is a three-body decay and hence suppressed by the propagator through $1/m_{W}^{2}$, it completely dominates over the two-body decay mode proportional to $|y_{N}|^{2}$, which can have a Branching Ratio (BR) of $\mathcal{O}(10^{-5})$.
However, 
the fact that $\tilde{\chi}^{\pm}\rightarrow \ell^{\pm}\tilde{N}$ is a two-body decay fixes the kinematics in the rest frame of the chargino. At the LHC this is not much of a help as the relationship between the laboratory frame and the rest frame of the chargino is unknown. At electron-positron colliders, instead,  the situation is different: if the collision energy is chosen so that the charginos are produced almost at rest, the energy of the charged lepton will be nearly fixed in the lab frame and this can be used in the event selection.

We therefore consider the process $e^+ e^-\to \gamma^{*} /Z^{*} \to \tilde{\chi}^+\tilde{\chi}^-$ with one of the charginos decaying to a lepton and a sneutrino and the other  decaying to a neutralino and a virtual $W^\pm$ leading to a soft lepton plus MET or to hadrons. If the neutralino decays invisibly through the neutrino Yukawa couplings, {\it i.e.}, via $\tilde{\chi}^{0}\rightarrow \tilde{N}\nu$, the final state with a single hard lepton and MET will get a too large background from $W^\pm$ bosons. If, however, the right-handed neutrino and sneutrino are light enough, the neutralino can decay via its singlino component through $\tilde{\chi}^{0}\rightarrow\tilde{N}N$ with the right-handed neutrino decaying subsequently to a lepton and two jets. The Feynman diagram of this process is shown in Figure \ref{fig:signal}.
\begin{figure}[h]
    \centering
    \begin{fmffile}{signal}
  \begin{fmfgraph*}(300,150)
    \fmfset{arrow_len}{3.5mm}
    \fmfstraight
    \fmfleft{i1,i2,i3,i4,i5}
    \fmfright{o1,o2,o3,o4,o5,o6,o7}
    \fmf{fermion,tension=3}{i4,v1,i2}
    \fmf{boson,label={$\gamma/Z$},tension=4}{v2,v1}
    \fmf{phantom,tension=1}{o2,v2,o6}
    \fmflabel{$e^+$}{i2}
    \fmflabel{$e^-$}{i4}
    \fmffreeze
    \fmf{fermion,tension=0.5}{v2,v3}
    \fmf{fermion,tension=0.5}{v2,v4}
    \fmf{phantom,label={$\Tilde{\chi}^+$},tension=0,label.side=right}{v2,v4}
    \fmf{phantom,label={$\Tilde{\chi}^-$},tension=0,label.side=left}{v2,v3}
    \fmf{phantom,tension=1}{o1,v4,o4}
    \fmf{phantom,tension=1}{o1,v5,o3}
    \fmf{phantom,tension=1}{o6,v3,o7}
    \fmf{phantom,tension=1}{i1,v4,v3,i5}
    \fmffreeze
    \fmf{fermion}{v3,o7}
    \fmf{scalar}{v3,o6}
    \fmflabel{$l^-$}{o7}
    \fmflabel{$\Tilde{N}$}{o6}
    \fmf{fermion,tension=2,label={$\Tilde{\chi}^0$},label.side=right}{v4,v5}
    \fmf{boson,tension=2}{v4,o5}
    \fmflabel{$W^+$}{o5}
    \fmf{fermion,tension=1.4,label={$N$},label.side=right}{v5,v6}
    \fmf{scalar,tension=0.5}{v5,o4}
    \fmflabel{$\Tilde{N}$}{o4}
    \fmf{fermion}{v6,o1}
    \fmf{fermion}{v6,o3}
    \fmf{fermion}{v6,o2}
    \fmflabel{$j$}{o3}
    \fmflabel{$j$}{o2}
    \fmflabel{$l$}{o1}
    \fmfshift{60 right}{o1,o2,o3}
    \fmfshift{40 left}{v5}
    \fmfshift{30 right}{o4}
  \end{fmfgraph*}
\end{fmffile}
    \caption{An example of a full process leading to the `dijet + dilepton + $\slashed{E}_T$' signature.\label{fig:signal}}

\end{figure}
Requiring a lepton and two jets with an invariant mass correspoding to the right-handed neutrino mass will be enough to get rid of the backgrounds, as we shall see. Finally, notice that the existence of right-handed neutrinos can be established already at the LHC \cite{Cerdeno:2013oya}.

Thus we can get the aforementioned signature: `dijet + dilepton + MET'. The dilepton signature should emerge in both same-sign and opposite-sign dileptons due to the Majorana nature of the right-handed neutrino. The latter will have a smaller background from SM processes. The major SM background to this final state comes from the following processes.

\begin{itemize}
    \item $W^+W^-Z$ production in the case where one $W^\pm$ boson decays into two jets and the other to a lepton and neutrino while the $Z$ boson gives two leptons, one of which is missed by the detector.
    \item $ZZZ$ production, where one $Z$ boson decays leptonically, the second to neutrinos and the third creates the two jets.
    \item $t\bar{t}$ production, where the top (anti)quarks decay to a $W^\pm$ boson and a $b$-quark, when one lepton originates from the $W^\pm$ boson and another lepton from a $B$-meson.
\end{itemize}
We now proceed to describe how we performed our Monte Carlo (MC) analysis.

\subsection{Event simulation}

We prepared the model files with the Mathematica package \textsc{Sarah} v4.14 \cite{Staub:2015kfa,Vicente:2015zba}, which creates the source code for \textsc{SPheno} v4.0.3 \cite{Porod:2003um,Porod:2011nf} to generate the mass spectrum. We simulate collider events with \textsc{MadGraph5} v2.8.2 \cite{Alwall:2011uj}. We use \textsc{Pythia} v8.2 \cite{Sjostrand:2014zea} for parton showering and simulate the detector response with \textsc{Delphes3} \cite{deFavereau:2013fsa}, where we use the DSiD card \cite{Potter:2016pgp} to simulate the detector at the future International Linear Collider (ILC) (see https://linearcollider.org/). We use \textsc{MadDM} v.3.0 \cite{Ambrogi:2018jqj} to check that our Benchmark Points (BPs) satisfy the constraints from the relic density and direct detection experiments. Lastly, we use \textsc{MadAnalysis5} v1.8 \cite{Conte:2012fm} to implement the cuts.

We prepared a number of BPs, which could be probed at the $\sqrt{s}=500$~GeV phase of the ILC. As the integrated luminosity we use $4000$~fb$^{-1}$, which could represent the total integrated luminosity after the luminosity upgrade of the ILC \cite{Barklow:2015tja}. We select the charginos to be slightly lighter than $250$~GeV and the right-handed neutrino and sneutrino so light that $\tilde{\chi}^{0}\rightarrow \tilde{N}N$ is kinematically allowed. We show the spectra and relevant Yukawa couplings of our BPs in Table  \ref{tb:benchmarks}. We checked with \textsc{MadDM} v3.0 \cite{Ambrogi:2018jqj} that  the BPs are acceptable with respect to constraints from the relic density and direct detetion experiments. Regarding the relic density we only imposed the upper limit $\Omega h^{2}\leq 0.12$.

\begin{table}
\begin{center}
\begin{tabular}{|c|c|c|c|}
\hline
 & BP1 & BP2 & BP3\\
\hline
$m(\tilde{\chi}_{1}^{\pm})$ (GeV) & 239.3 & 234.8 & 233.3\\
$m(\tilde{\chi}_{1}^{0})$ (GeV) & 233.3 & 228.7 & 227.3\\
$m(\tilde{N}_{1})$ (GeV) & 130.6 & 127.9 & 127.4\\
$m(N_{1})$ (GeV) & 101.7 & 90.5 & 88.6\\
$\mathrm{BR}(N\rightarrow \ell jj)$ & $60\%$ & $68\%$ & $68\%$ \\
$\mathrm{BR}(W^{*}\rightarrow \mathrm{leptons})$ & $28\%$ & $28\%$ & $28\%$\\
$\tan \beta$ & $2.3$ & $2.4$ & $2.1$\\
$y^{\nu}_{1j}$, $y^{\nu}_{2j}$ ($10^{-7}$) & $5.3$, $3.5$ & $6.1$, $4.7$  & $5.3$, $4.0$\\
\hline
\end{tabular}
\end{center}
\caption{Mass spectra and the most important parameters of our BPs. The neutrino Yukawa couplings are given for the flavour of the lightest right-handed sneutrino. The amplitude of our signal process will be proportional to these Yukawa couplings.\label{tb:benchmarks}}
\end{table}

The BPs cover a somewhat limited part of the parameter space, mainly because we wanted the charginos to be accessible at the ILC with $\sqrt{s}=500$~GeV and the decay $\tilde{\chi}_{1}^{0}\rightarrow \tilde{N}_{1}N_{1}$ needs to be kinematically open while simultaneously evading constraints from  searches for right-handed neutrinos. Such constraints are a lot stronger in the region $m_{N}<m_{W}$ \cite{CMS:2018iaf}. The scale of neutrino masses then forces the Yukawa couplings to be around $5\times 10^{-7}$.

As a preselection we require two same-sign leptons, two jets and veto against $b$-jets. For the $b$-tagger we use a working point, where the $b$-tagging efficiency is $70\%$ with a mistagging rate of $2\%$ for $c$-quark jets and $0.3\%$ for light quark (and gluon) jets. We summarise this preselection in Table \ref{tb:preselection}.

\begin{table}
\begin{center}
\begin{tabular}{|c|c|}
    \hline
    Number of leptons & $N(\ell)=2$ \\
    \hline
    Same-sign lepton pair & $N(\ell^+)$ or $N(\ell^-)$ = 2 \\
    \hline
    Number of jets & $N(j)=2$ \\
    \hline
    Veto on $b$-jets & $N(b)= 0$ \\
    \hline
\end{tabular}
\end{center}
\caption{The requirements for the final state topology. Here $\ell = e,\, \mu$.\label{tb:preselection}}
\end{table}

\begin{figure}
    \centering
    \includegraphics[scale=0.3]{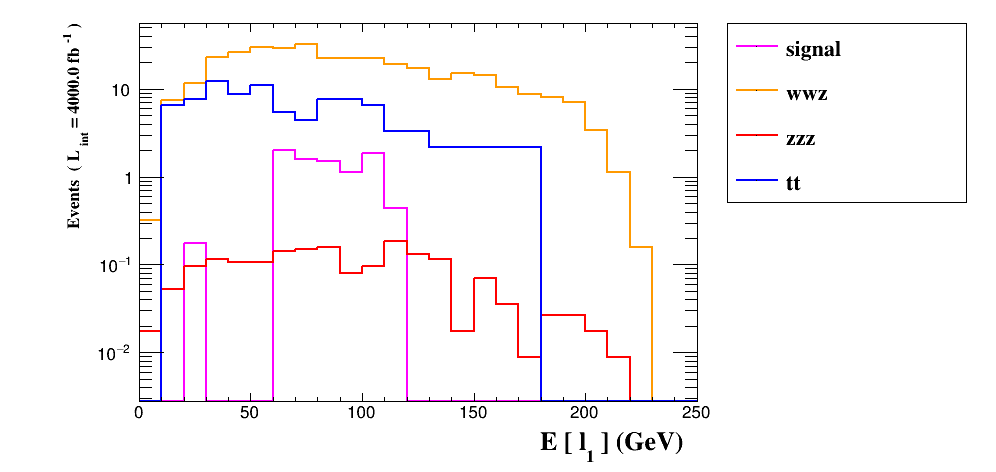}
    \caption{The energy of the leading lepton $\ell_1$ for our signal and different background components. The preselection of Table \ref{tb:preselection} has been imposed.}
    \label{fig:el1}
\end{figure}

We impose several cuts. The leading lepton $\ell_1$ arises from $\tilde{\chi}^{\pm} \to \tilde{N}\ell^{\pm}$, which is a two-body process. As long as the beam energy is not much larger than $2m_{\tilde{\chi}^{\pm}}$, the lepton energies in the lab frame are in a rather narrow range determined by the event kinematics as can be seen from Figure \ref{fig:el1}. We filter the signal by requiring $60\; \mathrm{GeV} <E(\ell_1)< 120$~GeV and add a requirement $p_{T}(\ell_{1})> 30$~GeV to ensure triggering by a single lepton + MET trigger.

The second lepton and the jets arise from $N\rightarrow \ell jj$ and the momenta of the decay products are limited by the mass of the right-handed neutrino. Hence, the momenta are constrained from above and imposing a veto against a second hard lepton with  $p_T(\ell_2)> 40$~GeV and against a hard jet with  $p_T(j_1)>70$~GeV leaves our signal untouched while rejects a reasonable fraction of the background.

As the hadronic activity in our signal events arises mainly from the hadronic decay of a $W^\pm$ boson emerging from the right-handed neutrino, we expect the total hadronic energy to be smaller than in the background events, especially $t\overline{t}$. Hence we require $H_{T}< 100$~GeV.

\begin{figure}
    \centering
    \includegraphics[scale=0.3]{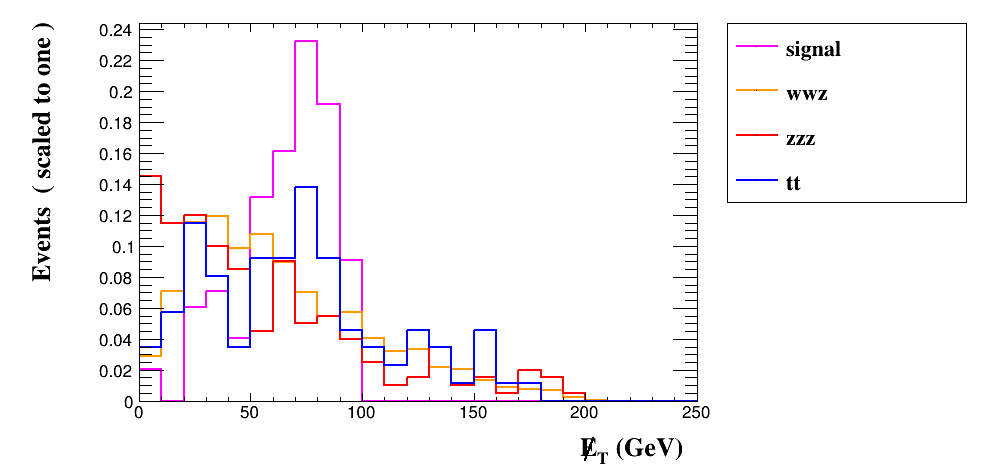}
    \caption{The distribution of missing transverse energy ($\slashed{E}_T$) for the signal and background components, again with the preselection of Table \ref{tb:preselection}. Here we have  normalised the distributions to unity.}
    \label{MET}
\end{figure}

\begin{figure}
    \centering
    \includegraphics[scale=0.3]{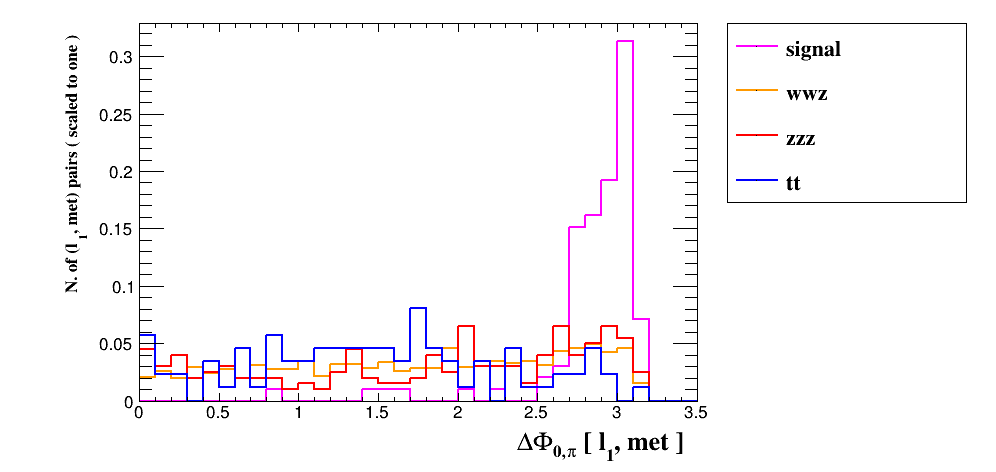}
    \caption{The distribution of the azimuthal angle between the leading lepton and missing transverse energy, the different distributions are normalised to unity.}
    \label{theta}
\end{figure}

\begin{figure}
    \centering
    \includegraphics[scale=0.3]{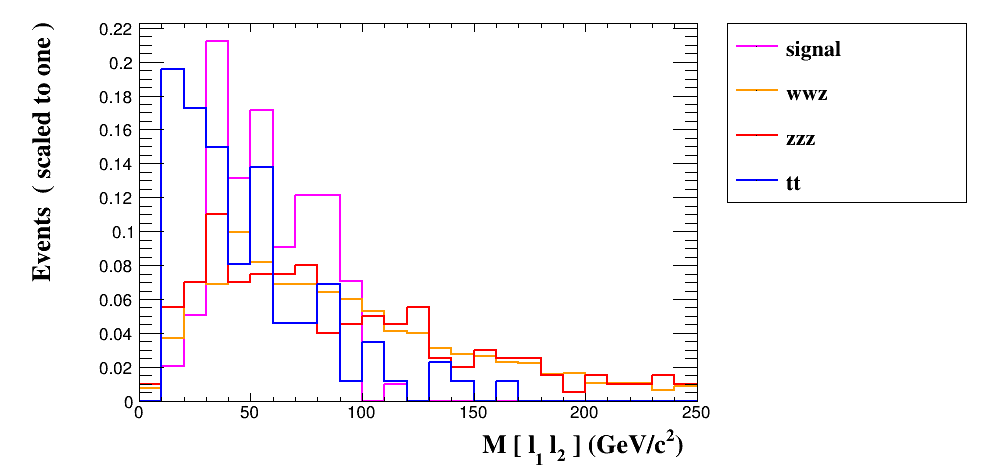}
    \caption{The distribution of invariant mass of the leading two leptons, the different distributions are normalised to unity.}
    \label{IMl1l2}
\end{figure}

The LSPs give rise to MET for  signal events. As we can see from Figure \ref{MET}, the distribution of $\slashed{E}_{T}$ for the signal is mostly in the interval $[50,100]$~GeV, hence, we select that interval. From Figure \ref{theta} we can see that for the signal the leading lepton is almost in the opposite direction compared to $\slashed{p}_{T}$ so we require the azimuthal angle between the leading lepton and missing transverse momentum to be greater than $2.5$ radians. We also impose the condition $M(\ell_{1}\ell_{2})<80$~GeV for the invariant mass of the  lepton pair, which rejects a fraction of the bosonic backgrounds as can be seen from Figure \ref{IMl1l2}.

Finally, we assume that the right-handed neutrino mass is known and require the invariant mass of the second lepton\footnote{The leading lepton arises from the two-body decay of the chargino, the one coming from the right-handed neutrino is always softer.} and the two leading jets to be close to the right-handed neutrino mass. We show the full list of cuts in  Table \ref{tb:cut} and the resulting cutflow for the signal  and background components in Table \ref{cutresult}.

\begin{table}
    \centering
    \begin{tabular}{|c|c|}
    \hline
    Transverse momentum & \\
 of leading lepton &  $ p_t(\ell_1) > 30$~GeV \\
    \hline
    Energy of leading lepton & 60 GeV $<E(\ell_1)<$ 120 GeV\\
    \hline
   Transverse momentum & \\ 
   of sub-leading lepton & $p_t(\ell_2)<$ 40 GeV \\ 
    \hline
    Transverse momentum & \\ 
    of leading jet &    $p_t(j_1)<$ 70 GeV \\
    \hline
    Total hadronic energy  & $H_T<$ 100 GeV \\
    \hline
    Missing transverse energy  & 50 GeV $<\slashed{E}_T <$ 100 GeV \\
    \hline
    Invariant mass of $\ell_1$ $\ell_2$ & $M(\ell_1\ell_2)<$ 80 GeV \\
    \hline
    Angle of leading lepton with & \\ 
    MET  & $\Delta\Phi_{0,\pi}>2.5$ \\
    \hline
    Invariant mass of two jets & \\ 
    and sub-leading lepton & $90$ GeV $<M(j_1j_2\ell_2)<110$ GeV \\
    \hline
\end{tabular}
\caption{The full set of cuts used in out MC analysis.}
\label{tb:cut}
\end{table}

\begin{table}[]
    \centering
\hspace*{-1.25truecm}
    \begin{tabular}{|c|c c c|c c c c|}
    \hline
    Cut & BP1 & BP2 & BP3 & $W^+W^-Z$ & $ZZZ$ & $t\bar t$ & Total background \\
    \hline
    Initial & 87.0 & 139 & 116 & 158999 & 4400 & 2193599 & 2356998 \\
    $b$-jet veto & 84.2 & 137 & 115 & 133754 & 2802 & 240648 & 377204 \\
    $N(\ell)$=2 & 38.8 & 54.9 & 42.0 & 11308 & 387 & 11454 & 23149 \\
    $N(\ell^+)=2$ or $N(\ell^-)=2$ & 17.8 & 26.0 & 20.6 & 792 & 6.07 &339 & 1137 \\
    $N(j)=2$ & 8.66 & 12.3 & 8.69 & 343 & 1.76 & 95.4 & 440\\
    \hline
   
    $p_T(j_1)<70$ GeV & 8.66 & 12.0 & 8.35 & 154.5 & 0.625 & 26.3 & 181.4 \\
    $p_T(\ell_1)>30$ GeV & 7.87 & 10.2 & 8.11 & 134.5 & 0.519 & 17.6 & 152.6 \\
    $p_T(\ell_2)<40$ GeV & 7.87 & 10.2 & 8.11 & 95.7 & 0.36 & 17.6 & 113.7 \\
    $H_T<100$ GeV & 7.87 & 10.2 & 8.00 & 76.5 & 0.24 & 11.0 & 87.7 \\
    
    $E(\ell_1)<120$ GeV & 7.87 & 10.2 & 8.00 & 55.5 & 0.176 & 7.68 & 63.4 \\
    $E(\ell_1)>60$ GeV & 7.87 & 9.33 & 7.65 & 36.6 & 0.123 & 5.48 & 42.2 \\
    $\Delta\Phi_{0,\pi}>$2.5 & 7.70 & 8.08 & 6.14 & 16.7 & 0.035 & 3.29 & 20.0 \\
    $\slashed{E}_{T}>50$ GeV & 6.82 & 7.38 & 4.98 & 9.70 & 0.026 & 2.19 & 11.9 \\
    $\slashed{E}_{T}<100$ GeV & 6.82 & 5.99 & 4.06 & 8.27 & 0.026 & 2.19 & 10.5 \\

    $M(\ell_1\ell_2)<80$ GeV & 5.60 & 5.71 & 3.94 & 4.77 & 0.018 & 1.10 & 5.89 \\
    $M(j_1j_2\ell_2)<110(100)$ GeV & 5.51 & 5.71 & 3.94 & 2.23(1.40) & 0.0088(0) & 1.10(1.10) & 3.34(2.50) \\
    $M(j_1j_2\ell_2)>90(80)$ GeV & 3.67 & 3.48 & 2.43 & 1.11(0.636) & 0.0088(0) & 0(0) & 1.1(0.64)\\
    \hline
    \end{tabular}
    \caption{The cutflow for the signal BPs and all background. The luminosity is 4000 fb$^{-1}$ and the energy is $\sqrt{s}=500$~GeV. The bracket stands for the cut and result corresponding to both BP2 and BP3. After these cuts, the significance for BP1 is $3.5\sigma$, for BP2 $4.4\sigma$ and for BP3 $3.0\sigma$.}
    \label{cutresult}
\end{table}

Overall we may see that we are able to see a significant excess above the expected SM background, showing evidence of the neutrino mass generation mechanism, but a full discovery would need a higher integrated luminosity. Nevertheless, even if the excesses are not statistically significant enough to claim the discovery of the two-body decay of the chargino, some bounds on the neutrino Yukawa couplings can be inferred.

\section{Estimating neutrino Yukawa couplings}

The coupling between the right-handed sneutrino, charged lepton and lightest chargino is
\begin{equation}
\lambda_{\tilde{N}\ell^{+}\tilde{\chi}^{-}}=\frac{i}{\sqrt{2}}y^{\nu}_{ab}V_{12}\frac{1+\gamma_{5}}{2},
\end{equation}
where $a,b$ refer to neutrino flavours and $V_{12}$ gives the higgsino component of the lightest chargino. For our BPs, we have $|V_{12}|\simeq 1$.
This leads to the following decay width (neglecting the lepton mass):
\begin{equation}\label{eq:charginowidth1}
\Gamma(\tilde{\chi}^{\pm}\rightarrow \ell^{\pm}_{a}\tilde{N}_{b})=\frac{(m_{\tilde{\chi}}^{2}-m_{\tilde{N}}^{2})^{2}}{64\pi m_{\tilde{\chi}}^{3}}|y^{\nu}_{ab}|^{2}|V_{12}|^{2}.
\end{equation}
The majority of charginos decay via $\tilde{\chi}^{\pm}\rightarrow \tilde{\chi}^{0}\ell^{\pm}\nu, \tilde{\chi}^{0}q\overline{q}^{\prime}$. This decay can in principle be mediated by several particles ($W^{\pm}$, $\tilde{\ell}^{\pm}$, $\tilde{\nu}$, $H^{\pm}$, $\tilde{q}$) and their contributions can interfere: explicit formulae for all possible contributions are given in \cite{Djouadi:2001fa}.

If we assume that the charged Higgs and all of the superpartners besides the right-handed  sneutrino and the higgsinos are heavy, the decay width of the chargino is calculable. In such a case the measurement of the BR of the rare chargino decay would give us an estimate of the neutrino Yukawa couplings through the computed full width and Eq.~(\ref{eq:charginowidth1}).
The overall chargino pair production cross section is readily calculated. The problem in estimating such a BR is that the observed number of sneutrino events after the full set of cuts does not represent the number of sneutrinos originally produced. Hence, we need to estimate how many sneutrinos are lost in the procedure.

If we assume that hadrons from the virtual $W^\pm$ decay cannot form a detectable jet, the only way the process of Figure  \ref{fig:signal} can produce two leptons and two jets is that the $W^\pm$ boson from the right-handed  neutrino decays hadronically. Then we can observe two leptons if either  the virtual $W^\pm$ decays hadronically (including hadronic taus) and we detect all leptons or the virtual $W^\pm$ decays leptonically and we miss one lepton. The only way to get same-sign dileptons is that one of the leptons detected arises from the right-handed neutrino. Hence, the probability of detecting two leptons that can lead to a same-sign dilepton signature is\footnote{If one would wish to follow the steps of the cutflow more precisely, one would have to add contributions, where the decay of the right-handed neutrino leads to two leptons (one being missed), which would be removed by the $N(j)=2$ requirement.}
\begin{eqnarray}
P(N(\ell)=2) & = & \epsilon(\ell_{1})\epsilon(\ell_{2})\times {\rm BR}(N\rightarrow \ell jj)\times {\rm BR}(W^{*}\rightarrow \mathrm{hadrons})\\ \nonumber
 & & +\epsilon(\ell_{1})\epsilon(\ell_{2})(1-\epsilon(\ell_{3}))\times {\rm BR}(N\rightarrow \ell jj)\times {\rm BR}(W^{*}\rightarrow \mathrm{leptons})\\ \nonumber
& & +\epsilon(\ell_{2})\epsilon(\ell_{3})(1-\epsilon(\ell_{1}))\times {\rm BR}(N\rightarrow \ell jj)\times {\rm BR}(W^{*}\rightarrow \mathrm{leptons}),\label{eq:lepteff}
\end{eqnarray}
where $\epsilon(\ell)$ is the lepton identification efficiency averaged over the detector. The probability of having two same-sign leptons is $50\%$ due to the Majorana nature of the right-handed neutrino. The probability of having two jets is $\epsilon(j)^{2}$, where $\epsilon(j)$ is the single jet identification efficiency, while the $b$-veto gives us a factor of $(1-a)^{2}$, where $a$ is the average mistagging rate.

We try to mimic the actual process of determining the selection efficiencies by basing the efficiency estimates on simulated data other than our primary process. We estimated the efficiencies of leading lepton and jet identification based on $e^{+}e^{-}\rightarrow W^{+}W^{-}$ data, which gives average values of $\epsilon(\ell_{1})=0.89 \pm 0.02$ and $\epsilon(j) = 0.83\pm 0.02$. The identification efficiency for the sub-leading lepton is lower. We estimated it from $e^{+}e^{-}\rightarrow \tilde{\chi}^{+}\tilde{\chi}^{-}$ data, where requiring at least one lepton, four jets and a fully reconstructed right-handed neutrino gives us a sample with over $95\%$ purity\footnote{Quite often the right-handed neutrino decay leads to a secondary vertex that can help reduce the backgrounds even further.}. From the numbers of single and dilepton events one may estimate the identification efficiency for the sub-leading lepton. This leads to estimate $\epsilon(\ell_{2})=0.50/0.46/0.45$ for BP1/BP2/BP3. The $b$-mistag rate is approximately $a=0.008$ as one in four jets arising from a hadronically decaying $W^\pm$ is a $c$-jet.
Overall this leads to a selection efficiency of $0.082/0.085/0.083$ for BP1/BP2/BP3 for the basic event selection of Table \ref{tb:preselection}. The systematic error of these efficiencies is $\mathcal{O}(10\%)$.

We then turn to the cut efficiencies. The cuts on the leading lepton transverse momentum and energy basically remove the cases where the leading lepton has not been identified, \textit{i.e} the third line of Eq. (\ref{eq:lepteff}). This gives a selection efficiency of $98\%$ for all of the benchmark points.
Some of the further cutting efficiencies can also be estimated in a data-driven way from chargino pair data. If we look at events with two leptons and four jets, where we are able to reconstruct the right-handed neutrinos, we may derive estimates of some kinematical distributions. Specifically, the sub-leading lepton and all hadronic activity of our signal events arise from the decays of the right-handed neutrino, so the efficiencies of the cuts $p_{T}(j_{1})<70$~GeV and $p_{T}(\ell_{2})< 40$~GeV can be directly inferred from data and the efficiency of requiring 
$H_{T}< 100$~GeV can be estimated by looking at the distribution of $H_{T}$ for events with two right-handed neutrinos.

For the cut on the second lepton momentum, the data-driven estimate for the cutting efficiency is $100\%$ for all BPs. For the leading jet momentum the efficiency is $100\%$ for BP1 and $98\%$ for BP2 and BP3. The total hadronic transverse momentum distribution would give an average of $67$~GeV with a $10$~GeV standard deviation, so that, assuming a Gaussian distribution, the signal acceptance is higher than $99.9\%$.

The rest of the cutting efficiencies need to be estimated by simulation. To successfully simulate events, the spectrum needs to be known. The chargino mass is easily determined from the onset energy of the chargino pair production. The problem is that the mass of the sneutrino is not too easy to determine from data. For instance, mono-$X$ signatures, which could give the LSP mass through the endpoint of the visible spectrum at $e^{+}e^{-}$ colliders \cite{Acciarri:1998hb,Abdallah:2003np} cannot be used in the case of a right-handed sneutrino LSP as they can be produced only via the Higgs portal or SUSY cascades. Eventually, the LSP mass can be measured from the endpoint of the lepton energy spectrum (which is $\sqrt{s}/2-m_{\tilde{N}}$) from the rare decay mode, but the number of events is so small that this may not provide a useful bound. Another limit for the LSP mass is naturally $m_{\tilde{N}}< m_{\tilde{\chi}}-m_{N}$, which can be improved with the $\slashed{E}_{T}$ distribution of the chargino pair events.

Overall we expect that the simulation should be able to determine the cutting efficiencies of the remaining cuts with a reasonable accuracy, perhaps with a $20\%$ error or so. In such a case, there would be an overall systematic uncertainty in the initial number of charginos decaying to the two-body state of about $30\%$. However, the statistical error would be larger, in the range of $50$--$70\%$ depending on the actual number of observed events. This would lead to an overall error in the range $55$--$75\%$ for the  BR$(\tilde{\chi}^{\pm}\rightarrow \ell^{\pm}\tilde{N})$. Since the latter  is proportional to the Yukawa coupling squared, this would give an error of about $25$--$35\%$ for the determination of the Yukawa couplings.

\section{Conclusions}

The right-handed sneutrino is an additional candidate to be the LSP in the NMSSM with right-handed neutrinos. If such a sneutrino is the LSP, the higgsinos need to  also be at the EW scale and potentially could be produced at the LHC and future $e^{+}e^{-}$ colliders. However, 
when the sneutrino is the LSP, we do not have the advantage of visible sneutrino decays as a window to neutrino physics. However, the charged higgsino has a small chance of decaying into a charged lepton and the sneutrino LSP, this decay being determined by the tiny neutrino Yukawa couplings.

In this paper, 
we have shown that, if the right-handed neutrino is so light that the neutral higgsino will decay via $\tilde{\chi}^{0}\rightarrow N\tilde{N}$, this additional handle will let us find the rare two-body decay of the chargino at the ILC,  $\tilde{\chi}^-\to l^-\tilde N$, while at the LHC such a discovery is impossible as the boost between the laboratory frame and the CM frame is unknown. This rare two-body decay would allow us to estimate the size of the neutrino Yukawa couplings but, even with the luminosity upgrade of the ILC, the measurement will be statistically limited and give at best an accuracy of $25\%$. Even this limited accuracy should be enough, though, to test the consistency of the seesaw model, \textit{i.e.} that the neutrino masses are generated by the Type-I seesaw and not by some more extended seesaw model like the inverse or linear ones, where the couplings could be orders of magnitude larger.

Clearly, the requirement introduced by the CM energy of the ILC being fixed at discrete values all below the TeV scale, as opposed to the LHC case where, at the partonic level, $\sqrt{\hat{s}}$ can be well beyond it, limits the region of NMSSMr parameter space that can be accessed. However, within this part of the parameter space we have then drawn representative BPs which show that we can see a signal from Yukawa couplings smaller than $10^{-6}$. 

\section*{Acknowledgments} SM  is  financed  in part through the NExT Institute and the STFC consolidated Grant No.  ST/L000296/1.  HW acknowledges financial  support  from  the  Finnish  Academy  of  Sciences and Letters and STFC Rutherford International Fellowship scheme (funded  through  the  MSCA-COFUND-FP  Grant No.   665593).   The  authors  acknowledge  the  use  of  the IRIDIS High Performance Computing Facility and associated support services at the University of Southampton, in the completion of this work.

\end{document}